\documentstyle[twoside,fleqn,espcrc2,epsfig]{article}

\newcommand{\AmS}{{\protect\the\textfont2
  A\kern-.1667em\lower.5ex\hbox{M}\kern-.125emS}}

\title{X-ray Imaging Using a Hybrid Photon Counting GaAs Pixel Detector}

\protect \author{C. Schwarz$^{a}$%
\thanks{Further author information: Tel: +49 761 203 5935, Fax: +49 761 203 5931, 
Email: Christoph.Schwarz@physik.uni-freiburg.de}, M. Campbell$^b$, R. Goeppert$^a$, E.H.M. Heijne$^b$
 J.Ludwig$^a$, G. Meddeler$^d$, B. Mikulec$^b$, E.~Pernigotti$^c$, M.Rogalla$^a$,
 K. Runge$^a$, A. S\"oldner-Rembold$^a$,
K.M. Smith$^e$, W. Snoeys$^b$, J. Watt$^e$ \\[2ex]
  $^a$ University of Freiburg, Hermann-Herder-Str. 3, 79104 Freiburg, Germany \\
 $^b$ CERN, Geneva, Switzerland \\
 $^c $ University and INFN of Pisa, Italy\\
 $^d$  Nikhef, Amsterdam, The Netherlands \\
 $^e $ University of Glasgow, Scotland}

\begin{document}

\begin{abstract}
The performance of hybrid GaAs pixel detectors as X-ray imaging sensors were investigated at
room temperature. These hybrids consist of 300~$\mu$m thick
GaAs pixel detectors, flip-chip bonded to a CMOS Single Photon Counting Chip (PCC). 
This chip consists of a matrix of 64 x 64 identical square pixels (170 $\mu $m x 170 $\mu $m)
and covers a total area of 1.2 cm$^2$. The electronics in each cell comprises a preamplifier, 
a discriminator with a 3-bit threshold adjust and a 15-bit counter. The detector 
is realized by an array of Schottky diodes processed on semi-insulating LEC-GaAs bulk material.
An IV-charcteristic and a detector bias voltage scan showed that the detector can be operated 
with voltages around 200~V. Images of various objects were taken by using 
a standard X-ray tube for dental diagnostics. The signal to noise ratio (SNR) was also determined.\\
The applications of these imaging systems range from medical applications
like digital mammography or dental X-ray diagnostics to non destructive
material testing (NDT). Because of the separation of detector and readout
chip, different materials can be investigated and compared.



\end{abstract}

\maketitle

\section{INTRODUCTION}

The most widely used detection medium for medical X-ray imaging is still photographic
film. In the last few years, also digital X-ray imaging systems have been playing an increasing role.
The main advantages of digital sensors in comparison to film systems are the higher sensitivity 
due to better absorption (this implies lower dose for the patient), the avoidance of time and 
material consuming chemical processing and the benefits of digital data handling like the
possibility to apply software image processing tools to analyze the image. 

\setcounter{footnote}{0}

The digital X-ray systems which are commercially available since a few years \cite{Weland}, mainly consist 
of silicon charge coupled devices (CCDs), with or without a scintillator conversion layer. 
Incident photons create electron hole pairs which are accumulated in potential wells formed by 
the electrodes of the CCD. These potential wells are located very close to the surface of the CCD. 
In contrast to visible light which can be absorbed very well in this thin region, the absorption
of X-rays is much less efficient due to higher photon energy. To increase the absorption, the CCD is 
often covered with a scintillator layer. This concept has the disadvantage of decreasing image resolution
and contrast because of scattering of conversion photons within the scintillator.

Another concept is given by hybrid pixel assemblies, which consist of a detector and a readout chip being
connected together by a flip-chip process. Different developments have been sucessfully made especially
for high energy physics and recently medical applications. A big advantage of the hybrid solution compared to 
monolithic devices like a CCD is the fact, that both chips can be optimized separately. While for 
the readout circuit the well known silicon CMOS technology is preferred, materials with an enhanced 
absorption efficiency for X-rays in the energy range of 10-70 keV such as GaAs or CdTe can be used.

A new step in the readout electronics is made by using a single photon counting technique instead of 
an integrating method. This implies a faster read-out, low noise and a higher dynamic range. In this work, 
detectors processed on semi-insulating LEC-GaAs (SI-GaAs) bulk material and bump-bonded to the Photon 
Counting Chip (PCC) \cite{Camp} were used.

\begin{figure}[tb]
\mbox{
          \epsfxsize=7.5cm
          \epsffile{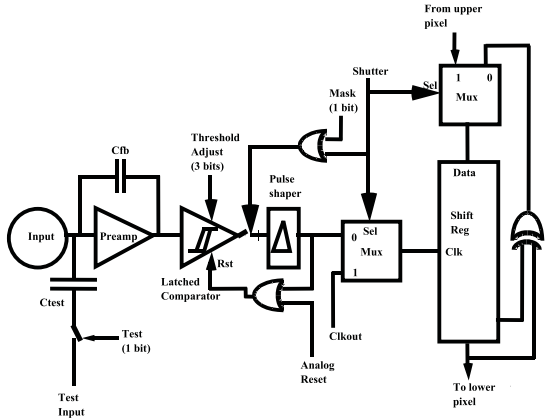}
           }
\vspace{-1cm}
\caption{Block diagram of the pixel cell.}
\label{fig:blockdiag}
\end{figure}

\vspace*{0.5cm}

\section{READOUT ELECTRONICS}

The PCC is a further development of the LHC1/Omega3 \cite{Heijne} chip, used in high energy physics, 
towards medical applications. It consists of a matrix of 64 x 64 matrix of identical square pixels
(170 $\mu $m x 170 $\mu $m) and covers a total sensitive area of 1.2 cm$^2$. 
The electronics in each cell comprises a preamplifier with a leakage current compensation
up to 10~nA/pixel, an externally adjustable comparator with a 3-bit fine tuning for each
pixel, a short delay line which feeds back to the latched comparator to produce a pulse 
and a 15-bit pseudo-random counter. The input of the preamplifier is connected via a bump-bond
to one of the detector pixels or can receive alternatively test signals from an external pulse 
generator via a test capacitance. When the shutter signal is low the pulse coming from the delay line 
becomes the clock of the counter. When the shutter is high, the feedback loop is broken 
and an external clock can be used to shift out the data in a serial fashion. The maximum readout
frequency is 10~MHz. There are two more fully static flip-flops to mask noisy pixel and to
enable electrical testing.

A summary of electrical measurements of the PCC before bump-bonding is given in 
table~\ref{tab:electric}.

\begin{table}[tb]
\setlength{\tabcolsep}{1.5pc}
\newlength{\digitwidth} \settowidth{\digitwidth}{\rm 0}
\catcode`?=\active \def?{\kern\digitwidth}
\caption{Summary of electrical measurements of the 
PCC~\protect\cite{Bisog}.}
\label{tab:electric}
\begin{tabular}[t]{p{35mm} p{16mm}}
\hline
Minimum threshold                       & 1400~e$^-$  \\
\hline
Threshold linearity                     & 1400~e$^-$ to 7000~e$^-$ \\
\hline
Threshold variation \newline (no adjust)        & 350~e$^-$ \\
\hline
Threshold variation \newline (adjust)           & 80~e$^-$  \\
\hline
Noise                                   & 170~e$^-$ \\
\hline
Maximum input signal                    & $>$~80000~e$^-$     \\
\hline
Maximum counting \newline rate                  & 2~MHz \\

\hline
\end{tabular}
\end{table}

\vspace*{0.5cm}

\section{DETECTOR - MATERIAL AND \\ FABRICATION}

The detectors were fabricated in the Freiburg cleanroom facility on semi-insulating GaAs bulk 
material from FCM Freiberg, Germany. This material has typically a resistivity of 
10$^{7}\Omega$~cm. It has been shown that this type of GaAs has very good properties as a material 
for radiation detectors in high energy physics \cite{Rogal} and previously medical applications. 
The wafers were first lapped down from 650~$\mu$m to 300~$\mu$m and implanted on the backside with 
oxygen ($3\cdot 10^{13}$cm$^{-2}$ at 190~keV) to avoid backside firing. The Schottky contacts 
were processed on both sides by layers of Ti, Pl, Au and Ni. The front side is structured by
photolithographic processes into a matrix of small pixels (gap: 10~$\mu$m or 20~$\mu$m) of the 
same dimension as the readout electronics. 
The bond pads have a diameter of 20~$\mu$m, the passivation was made with a layer of Si$_3$N$_4$.
The so called underbump metallization for the flip-chip process was another layer of Au with an 
overlap of 2~$\mu$m. 

\begin{figure}[tb]
\mbox{
          \epsfxsize=7.5cm
          \epsffile{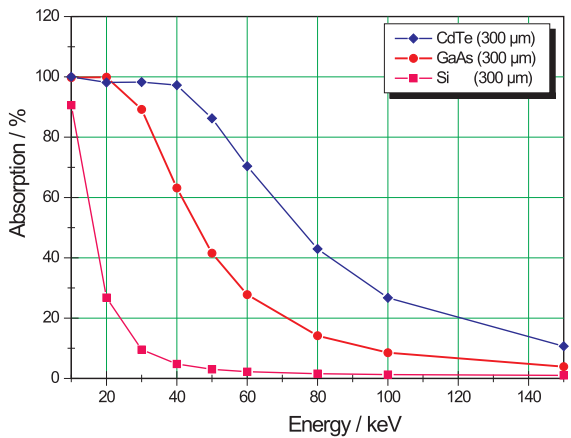}
           }
\vspace{-1cm}
\caption{Absorption of X-rays in different materials.}
\label{fig:material}
\end{figure}

In figure \ref{fig:material} the absorption probability of X-rays in the energy range 10 to 150~keV  
for Si, GaAs and CdTe, each 300~$\mu$m thick, is plotted. It can be seen that the absorption of GaAs and 
CdTe in the interesting energy range 20 to 70~keV is much higher than that of Si: for example at 30~keV,
which is the peak of the X-ray spectra for 70~kV tube voltage, the detection efficiency in Si is only 10\%,
in contrast to GaAs with nearly 90\%. CdTe performs even better, but until now there are difficulties in terms
of homogeneity and processing. 

\begin{figure}[tb]
\mbox{
          \epsfxsize=7.5cm
          \epsffile{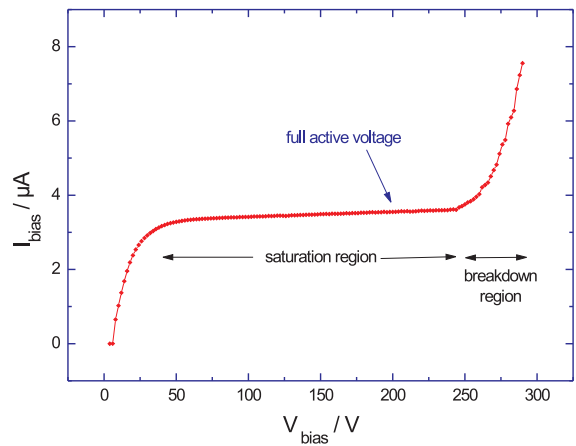}
           }
\vspace{-1.1cm}
\caption{IV-characteristic of a flip-chip bonded assembly.}
\label{fig:iv_charac}
\end{figure}

To determine the suitable reverse bias voltage settings of the detector, a IV-characteristic was taken 
with one assembly after the flip-chip bonding. A diode characteristic is expected. 
In figure~\ref{fig:iv_charac} the leakage current in $\mu$A which flows into the detector is plotted as a 
function of the reverse bias voltage.

The characteristic has three distinct regions: 
\begin{itemize}
\vspace{-1ex}

\setlength\itemsep{0ex}
\setlength\parsep{0ex}

        \item a region where the leakage current increases linear with reverse bias voltage to reach a plateau,
        \item a saturation area in which the leakage current is approximately independent of the bias voltage,
        \item a region where the current increases again with the applied voltage (soft breakdown region).
\end{itemize}

The soft breakdown is obtained due to the implantation of the backside of the wafer. The leakage current 
density measures 27~nA/mm$^2$.

\vspace*{0.5cm}

\section{IMAGING PROPERTIES}

To determine the imaging properties of the detector assembly we use a standard X-ray tube for dental 
diagnostics\footnote{Supplier: Siemens Type:Heliodent MD.}.
Measurements using radioactive sources have been sucessfully done by other groups \cite{Bisog}.

\begin{figure}[tb]
\mbox{
          \epsfxsize=7.5cm
          \epsffile{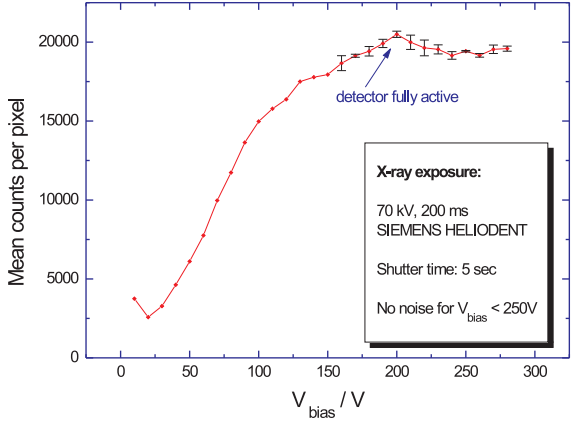}
           }
\vspace{-1cm}
\caption{Bias voltage scan of a flip-chip bonded assembly: Mean counts per pixel.}
\label{fig:biasscan}
\end{figure}

\begin{figure}[tb]
\mbox{
          \epsfxsize=7.4cm
          \epsffile{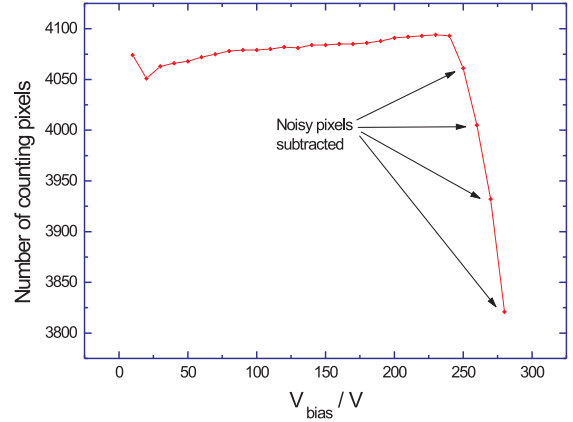}
           }
\vspace{-1cm}
\caption{Bias voltage scan of a flip-chip bonded assembly: Number of counting pixels.}
\label{fig:noise}
\end{figure}

In a first measurement the assembly was exposed to a 200~ms long, 70~kV X-ray pulse, and the mean counts 
per pixel for increasing reverse bias voltage were determined. We found that the mean counts per pixel
reach a plateau at around 200~V (figure~\ref{fig:biasscan}) and that there are almost no noisy pixels up
to 250~V (figure~\ref{fig:noise}). It should be mentioned that the bias settings of the readout chip 
were determined before, using an external pulse generator. The mean threshold
of the pixels after adjustment was calculated to be 3794~e$^-$.

First images were taken from a 10~mm long M2 steel screw, placed on the back of the detector in a distance of
20~cm from the X-ray tube. The tube voltage was set to 60~kV, the exposure time to 50~ms. These are nearly the
minimum settings of the tube. 

\begin{figure}[tb]
\mbox{
          \epsfxsize=7.5cm
          \epsffile{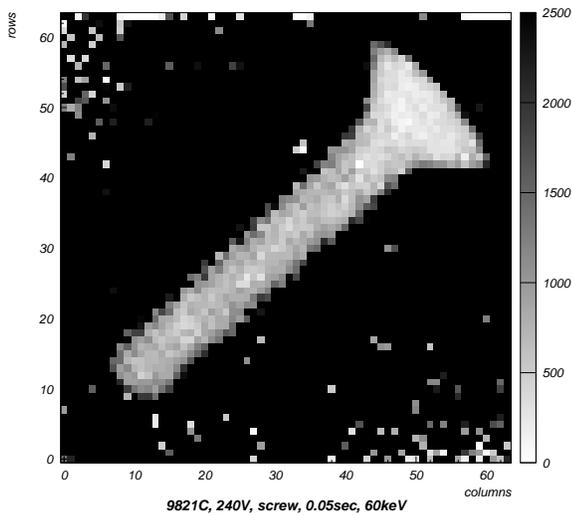}
           }
\vspace{-1cm}
\caption{X-ray image of a M2 steel screw, 0.05~s at 60~kV.}
\label{fig:m2_screw}
\end{figure}

In figure~\ref{fig:m2_screw} the raw data (the number of counts for each pixel) is plotted for the whole
pixel matrix. The darker the pixel is plotted in the 8-bit greyscale, the higher the count rate and so the
number of photons being detected in this pixel. Pixels which are plotted black, have counted more than 2500.
All pixels are working, so the bump-bonding yield of this assembly seems to be nearly perfect. Nevertheless there
are some small inhomogeneities, which can be attributed to variations in sensitivity of the detector.

This non-uniform sensitivity is probably a characteristic of the used bulk material. In semi-insulating
LEC-GaAs, the deep donor arsenic antisite defect EL2 is normally used to compensate residual
impurities with a flat energy level and is responsible for the semi-insulating behaviour. Otherwise the
influence of flat acceptor concentrations like carbon would leave the material conducting.

It has been shown that these deep donors could limit the lifetime of charge carriers by acting as trapping
centres \cite{Rogal2}. Electron-hole pairs generated by incoming X-ray photons can be trapped on their way to
the readout electrodes, so that only a fraction of the generated charge is detected. This implies a reduced
charge collection efficiency (CCE).
 
The local inhomogeneities also reduce the signal to noise ratio, which is defined as follows:

\begin{equation}
 \mathrm{SNR}=\frac{\mathrm{signal}}{\mathrm{noise}}=\frac{\langle\mathrm{n}\rangle}{\sigma} \nonumber
\end{equation}

Here $\langle\mathrm{n}\rangle$ represents the mean number of counts per pixel in the region of interest and 
$\sigma$ is the standard deviation of the signal value. In case of photonic noise the SNR should
have a square root dependency on the mean count rate as expected by the poisson statistic. 
Depending on the bias voltage of the detector, the exposure time to the X-rays and the optical
density of the object and its spatial frequency, the SNR is not fixed. 
We obtained for the SNR a value of $4.1\pm0.1$ by taking a flood image, i.e. a uniform exposure 
of the whole detector, for 200~V bias voltage and a 100~ms long X-ray pulse at 70~kV tube voltage
without applying any corrections to the data.

It has been shown \cite{Irsig} that in the case of time independent inhomogeneities in detector sensitivity
an image correction method can be used to ameliorate the imaging properties. This method also increases
the SNR by decreasing the $\sigma$. Further investigations will show if this method is also suitable 
for our detector system. Another possibility to get a better homogeneity is given by the threshold adjust 
facility of the PCC. Instead of adjusting the individual pixel threshold with a pulse on the test capacitance like
it is done till now, an adjustment using the mean detector response to X-ray exposure could be carried out.

\begin{figure}[tb]
\mbox{
          \epsfxsize=6.4cm
          \epsffile{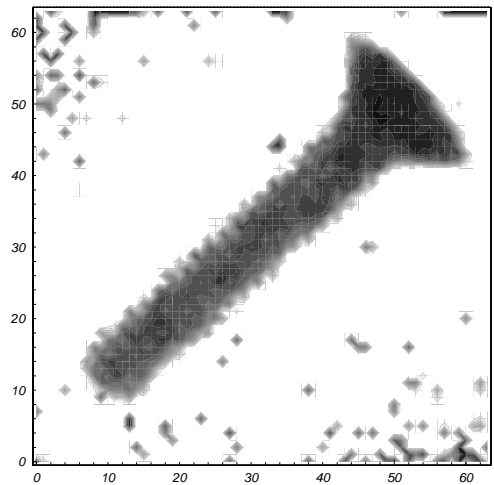}
           }
\vspace{-0.25cm}
\caption{Interpolated and inverted X-ray image of a M2 screw, 0.05~s at 60~kV.}
\label{fig:m2_screw_cont}
\end{figure}

To improve the image quality in a first step, the image was inverted and interpolated. This is shown in figure
\ref{fig:m2_screw_cont}. It should be mentioned that also the inner structure of the screw (thread, head)
can be recognized.

There are many ways to evaluate the quality of an image. The most common and suitable methods are
the contrast transfer function (CTF) and the modulation transfer function (MTF). The CTF describes 
the relative contrast response of an imaging system to a square wave modulation, the MTF the response 
to a sinusoidale one. They are both dependent on the spatial frequency whose unity is line 
pairs per mm (lp/mm). The Nyquist frequency which is defined by $N_y=1/(2\times \mathrm{pitch})$ 
measures 2.95~lp/mm for our detector system. Images of small slits down to the pixel size were sucessfully 
taken and the determination of the line spread function (LSF) and the corresponding MTF will be done soon.

\section{CONCLUSION AND \\ FUTURE WORK}

It has been shown that hybrid GaAs pixel detectors with photon counting electronics offer a 
promising alternative as digital X-ray imaging sensors. In this work SI-GaAs detectors, 
fabricated in Freiburg, were flip-chip bonded to 4096 Pixel Photon Counting Chips (PCC), developed at CERN. 
The leakage current density of the detector was determined by a IV-characteristic to 27~nA/mm$^2$, 
which is in accordance to the expectation. A detector bias voltage scan showed that a voltage 
around 200~V is enough to have the detector fully active. There are almost no noisy pixel for 
voltages below 250~V, the soft breakdown region of the detector.

Future work is given by investigations of the observed inhomogeneity in the taken X-ray images. If
they can be attributed to variations in sensitivity of the detector and are time independent,
an image correction methode can be developed and applied to the data. As a next step, characteristic 
quantities of an imaging system like the CTF and the MTF, will be determined and compared to other 
systems. 

\section{ACKNOWLEDGEMENTS}

This work was supported by the European Community under the Brite/Euram project XIMAGE (BE-1042).
The readout chip was developed as part of the Medipix project, carried out by
CERN, University of Freiburg, University of Glasgow and INFN-Pisa. We gratefully acknowledge
the contribution of G. Humpston of GEC-Marconi Materials Technology Ltd., Caswell, England for the bumb-bonding,
G. Magistrati of Laben S.p.A., Milano for the VME-based readout system and M. Conti and collaborators of
INFN-Napoli provided the readout software.

\end{document}